# Atomic scale characterization of nitrogen-doped graphite: Effects of the dopant nitrogen on the electronic structure of the surrounding carbon atoms

Takahiro Kondo[1][†], Simone Casolo[2][†], Tetsuya Suzuki[1], Taishi Shikano[1], Masataka Sakurai[1], Yoshihisa Harada[3,4], Makoto Saito[3][††], Masaharu Oshima[3,4], Mario Italo Trioni[5], Gian Franco Tantardini[2,5] and Junji Nakamura[1]*

[†] The authors contributed equally

* corresponding author: nakamura@ims.tsukuba.ac.jp

1 Faculty of Pure and Applied Sciences, University of Tsukuba, 1-1-1 Tennodai, Tsukuba, Ibaraki 305-8573, Japan

2 Dipartimento di Chimica Fisica ed Elettrochimica, Università degli Studi di Milano, via Golgi 19, 20133 Milan, Italy

3 Department of Applied Chemistry, The University of Tokyo, 7-3-1 Bunkyo-ku, Tokyo 113-8656, Japan

4 The University of Tokyo Synchrotron Radiation Research Organization, 7-3-1 Bunkyo-ku, Tokyo 113-8656, Japan

5 CNR National Research Council of Italy, ISTM, via Golgi 19, 20133 Milan, Italy

**Abstract:**

We report the local atomic and electronic structure of a nitrogen-doped graphite surface by scanning tunnelling microscopy, scanning tunnelling spectroscopy, X-ray photoelectron spectroscopy, and first-principles calculations. The nitrogen-doped graphite was prepared by nitrogen ion bombardment followed by thermal annealing. Two types of nitrogen species were identified at the atomic level: pyridinic-N (N bonded to two C nearest neighbours) and graphitic-N (N bonded to three C nearest neighbours). Distinct electronic states of localized $\pi$ states were found to appear in the occupied and unoccupied regions near the Fermi level at the carbon atoms around pyridinic-N and graphitic-N species, respectively. The origin of these states is discussed based on the experimental results and theoretical simulations.

[††]present address: Toyota Motor. Corporation, 1200, Mishuku, Susono, Shizuoka, 410-11 Japan

# I. Introduction

Nitrogen doping in graphitic materials such as graphene and carbon nanotubes has been reported to modify the physical and chemical properties [1-8]. Various applications of N-doped graphite (NG) materials have also been reported, such as in biosensors [9], fuel cells [10-15], capacitors [16], electronic devices [17-19], and spin filter devices [20, 21]. By X-ray photoelectron spectroscopy (XPS), the nitrogen species in NG materials [12, 13, 15, 19, 22-26] have been identified mainly as pyridinic-N (characterized by a binding energy of about 398.5 eV), pyrrolic-N (400.5 eV) and graphitic-N (or quaternary-N) (401.2 eV) [6, 27-29]. However, a detailed picture of the local electronic modifications in NG induced by each of N atoms has not been clarified.

The electronic structure near the Fermi level ($E_F$) of the graphite surface is known to be modified by the defect such as a point vacancy, where localized π electronic states are reported to form at the neighbouring C atoms and propagate anisotropically around the defect even to a few nm away from the defect due to the perturbation of the π-conjugated system of graphite [30]. It is thus expected that the nitrogen dopants in NG modify the electronic structure near $E_F$ and the modified electronic structure is different from that of the point vacancy defect because the nitrogen bonds to some of carbon atoms. It is also expected that the modified electronic structure is different depending on the types of C-N bonding configurations

Here, we report the atomic-scale characterization of a nitrogen-doped graphite surface by scanning tunnelling microscopy (STM), scanning tunnelling spectroscopy (STS), XPS and first-principles calculations based on the density functional theory (DFT). Two types of nitrogen species, graphitic-N and pyridinic-N, were identified at the atomic scale and effect of each nitrogen species on the local electronic structure of the surrounding carbon atoms is discussed based on the experimental results and theoretical simulations.

**II. Experimental**

Nitrogen-doped graphite was prepared *in situ* in all experiments reported here as follows. A fresh highly oriented pyrolytic graphite (ZYA-grade HOPG, Panasonic Inc.) sample was first cleaved in air using an adhesive tape and was placed in an ultrahigh vacuum (UHV) chamber. Subsequently, it was annealed at 900 ± 50 K. The sample was then bombarded by nitrogen ions of 200 eV (or 500 eV) at 300 K and normal incidence using a commercial ion gun (ANELVA, 5 kV ion gun). The total ion doses on the sample were set to be lower than $0.2 \times 10^{13}$ ions/cm$^2$, which were measured independently using a Faraday cup (for XPS experiment, we didn't measure the ion current). After the bombardment, the sample was annealed at 900 ± 50 K for 300 s. Based on the STM observations, the irradiated ions are found to form defects in the sample with high efficiency (see supplemental material, Fig. S1). Some of the nitrogen atoms were implanted into the bulk of the graphite, and as a result, the STM image showed peculiar characteristics, even after further cleavage in air followed by annealing in UHV. The sample temperature during heating was measured using an infrared thermometer.

XPS spectra were measured at 160 K in the UHV chamber with a base pressure of $3 \times 10^{-10}$ Torr at BL27SU in SPring-8 in Hyogo, Japan. The excitation photon energy used in this work was 700 eV. The X-ray excitation and photoelectron detection angles were set to 45° with respect to the sample normal. The total energy resolution was approximately 0.24 eV. Energy calibration was performed using the Au $4f_{7/2}$ photoemission at 84.0 eV.

STM and STS measurements were performed in the UHV chamber (UNISOKU, USM-1200) with a base pressure of $1 \times 10^{-10}$ Torr. A commercial STM controller (SPM-1000, RHK Technologies) was used. The sample was cooled to 5.3 K. The surface morphology and the surface electronic structure were then examined by STM and STS, respectively, with a PtIr tip (Pt:Ir = 8:2). STM images were recorded in the constant-current mode. STS measurements were carried out by measuring the differential conductance (*dI/dV*) by a lock-in detection technique using 1.0 kHz a.c. modulation of the sample bias voltage with an amplitude of 20 mV. *dI/dV* reflects features in the local density of states (LDOS) of the surface at the position of the STM tip, where *dI/dV* at negative and positive sample biases correspond to LDOS of occupied and unoccupied states, respectively, and 0 V corresponds to the Fermi level [31]. We used a small set current of 50- 200 pA to minimize the effect of the tip on the tunnelling spectrum. The STS spectra shown here represents the average of ten spectra measured at the same position.

**III. Theoretical**

The electronic structure and density of states (DOS) of the N-doped graphite sample were studied using DFT.

In theoretical simulations, an orthorhombic cell was preferred to a hexagonal one in order to avoid any (fictitious) band gap due to superlattice effects [32]. The supercell was made by graphene layers of 96 C atoms each, and defects were introduced in both a single layer or in a three layer cell, on top of which a 20 Å vacuum was used to avoid interaction between periodic replicas. The graphene layers were kept fixed at the experimental interlayer distance because of the well known failure of DFT in dealing with dispersion forces. After careful testing, we found that three graphene layers (for a total of 288 atoms) are sufficient to obtain converged properties.

Equilibrium geometries for N-doped graphite were first obtained using a plane wave density functional theory as implemented in the VASP code [33], by using a 500 eV energy cutoff and a $3\times3\times1$ $\Gamma$ centered k points mesh. The relaxation of the atomic position was stopped when the highest force acting on each ion was lower than 0.03 eVÅ$^{-1}$.

Core electrons, here considered as frozen, were included through projector augmented wave (PAW) potentials [34, 35] while a small (0.05 eV) Gaussian smearing was used to improve electronic convergence. Exchange and correlation was taken into account according to the gradient corrected Perdew-Burke-Ernzerhof (PBE) [36] functional. Later, atomic positions were optimized again with the SIESTA code [37] using a double-ζ plus polarization basis set, obtaining only negligible differences.

STM images and STS spectra were all simulated using SIESTA. Accurate local density of states were computed from the electronic ground state on a $100 \times 100 \times 1$ k-points mesh and as STS simulations, we show the DOS projected on $p_z$ orbitals of the topmost layer only, being the ones mainly probed by the STM tip. The Fermi energy was set to the Dirac point, because this is the limit for diluted defects. Simulated STM images were obtained within the Tersoff-Hamann approach [31] (constant current mode) for a doped single-layer graphene system that was found sufficient to give good agreement with the experiment. In this work, the spatial distribution of the selected electron densities (obtained from the wave function square modulus) are shown as isosurface plots from a top view.

**IV. Results and Discussion**

The XPS and STM measurements of nitrogen-doped graphite are described at first in sections IV-A. The local electronic structures of the surface near the Fermi level around the characterized dopant nitrogen species, pyridinic-N and graphitic-N, are then discussed in sections IV-B and IV-C, respectively, based on the experimental results and DFT calculations.

**A. XPS and STM measurements**

Fig. 1(a) shows N 1s XPS spectra of the nitrogen-doped graphite sample, where the results are shown for the surface bombarded with nitrogen ions at 200 eV before and after annealing at 900 K. The total amount of nitrogen with respect to that of carbon decreased from 3.1 to 2.7 at% after annealing, where the amounts were estimated by simply calculating the ratio between C 1s and N 1s peak areas and by considering the sensitivity factors, i.e. nitrogen atoms were assumed to be doped uniformly into deeper graphite layers. The decrease in the amount of nitrogen during annealing up to 900 K was possibly due to the desorption of $N_2$ or CN. Four N1s peak components at 398.5 eV, 399.9 eV, 401.1eV, and 403.2 eV have been assigned to pyridinic-N (N connected to two C), pyrrolic-N (N part of a pentagon ring connected to two C), graphitic-N (N connected to three C), and oxide-N species, respectively [6, 27-29]. Here, we note that the energy position for pyrrolic-N is known to be similar to that for cyanide-N (N connected to single C, 399.5 eV) and amine-N (N connected to single C and two H, 399.4 eV) [27], possibly also contributing to the same peak. In our samples, pyridinic-N and graphitic-N were the dominant components as shown in Fig. 1(a). The same two peak components have been reported as dominant for nitrogen-doped graphite or graphene prepared by the ion bombardment method, also for different collision energies (0.1–6.0 keV) and/or different annealing temperatures [23, 25, 26, 38]. However, the ratio of the peak intensities depends on the ion collision energy, flux of ions, and annealing temperature. In Fig. 1(a), graphitic-N is the largest component (40.1%) after nitrogen ion bombardment at 200 eV. After annealing to 900 K, the relative intensity of this peak increased (54.3%), suggesting that the other species were converted to graphitic-N. It has been reported that the relative amount of pyridinic-N is comparable with respect to that of graphitic-N when the sample is prepared with higher energies (a few keV) and more intense ion fluxes of nitrogen [23, 25, 26, 38]. This suggests that the graphitic-N rich surfaces can be obtained with small doses and low energies of nitrogen ion.

A typical STM image of the nitrogen-doped graphite surface is shown in Fig. 1(b). Two types of bright species were observed with an average diameter of about 4 nm: one was surrounded by a bright region (type A), and the other was surrounded by a dark region (type B). Because it is generally difficult to identify surface atom elements by STM alone, we have carried out the characterization by combining STS analysis and DFT calculations. We show here that type A and type B defects can be assigned to pyridinic-N species with a single-atom vacancy and graphitic-N species, respectively. These indeed correspond to the two dominant components observed in the XPS spectra of Fig. 1(a). The effect of each dopant N on the electronic structure near the Fermi level at the surrounding carbon atoms are discussed below. Note that in Figs. 1(c) and 1(d) some of the bright regions consist of the well-known superstructure of graphite formed by the standing wave of $\pi$ electrons around the graphite surface defects [39, 40]. In particular, this region does not spread isotropically around the defect but seems to propagate in three directions, as observed in the case of a single-atom vacancy defect on a graphite surface [30].

**B. Pyridinic-N**

The experimentally observed STM image and STS spectrum of a type-A defect are shown in Fig. 2. The STS spectrum shown in Fig. 2(b) was measured at the position indicated by the arrow in Fig. 2(a), and it consists of a large peak at about -370 mV and smaller peaks within a parabolic background. The spectrum is different from that of a single-atom vacancy with a single and large STS peak at [41] or just above the Fermi level [30, 42]. The STS spectrum of the type-A defect is again different from that of the single-atom vacancy in graphite in terms of a propagation feature [30]. That is, the modified electronic states of the single vacancy propagate anisotropically to three directions, while such an anisotropic propagation was not observed for the type-A defect as shown in Figs. S6-S8 in the supplemental material. To identify the defect species in Fig. 2(a), STM images and STS spectra were simulated for several types of structures having the pyridinic-N or the graphitic-N species because these species are the dominant species in our sample as shown by the XPS peak components in Fig. 1(a). After comparison with the results of different defect models, it was found that the simulated STM and STS features of pyridinic-N well reproduced those measured in experiments as shown in Fig. 2, That is, the propagation of the bright region in STM and the appearance of the STS peak at about -370 mV (Figs. 2(a) and 2(c)) are common between experiment and theory. The defect was thus assigned as the pyridinic-N.

From the DFT calculation, the large STS peak at -370 mV in Figs. 2(b) and 2(d) can be assigned to localized π states ($p_z$ orbitals) because of the localized DOS character (see supplemental material, Fig. S2). As shown in Fig. 3(a) the spatial propagation of the localized π states can be seen in the STM simulations at negative bias of -1.0 V. Most of the bright spots in Fig. 3(a) correspond to the LDOS for the STS peak at -370 mV (see supplemental material, Fig. S2) and are located on few C atoms, part of the same sublattice hosting the N atom.

To further study the electronic structure near the nitrogen atom, the DFT equilibrium geometry and charge density isosurfaces are examined as shown in Figs. 3(b) and 3(e). The isosurface stands for the spatial distribution of orbitals with a certain electron densities as described in Sec. III. First, pentagonal ring is found to be formed by the rearrangement of the two "edge" carbon atoms facing to the nitrogen atom, which is similar to the case of the C vacancy in graphene [43]. Second, the N atom site is imaged as dark in the simulated STM image at -1.0 V (Fig. 3(a)), even though nitrogen atom should bear one more electron compared to carbon. This is explained by the occupation of four electrons in the $sp^2$ coplanar orbitals of N with a lone pair as can be seen in Figs. 3(c) and 3(f) where the in-plane contribution to the electron density is represented by (blue-red) colours.

Here, we discuss the appearance of the localized states in the occupied region. The formation of the lone pair suggests that the N atom of the pyridinic defect is negatively charged. The negatively charged N is consistent with the chemical shift in N1s binding energy to lower energy for the pyridinic-N (398.5 eV) in the XPS spectrum compared to that for the graphitic-N (401.1 eV). If N is negatively charged, the surrounding carbon atoms would be charged positively because of the screening effect. The positive charge of carbon can explain a shift of the localized $\pi$ state of carbon from the Fermi level to the lower energy level. Following this picture, we suggest that the states corresponding to the STS peak in the occupied region near the Fermi level may act as Lewis base toward molecular species.

At positive bias, i.e. for the unoccupied region, the isosurface plot in Fig. 3(e) shows a nodal plane across the pentagon, suggesting an anti-bonding character of the states lying above the pentagon ring. A slight long range propagating feature can be observed in the simulated STM image of Fig. 3(d), suggesting the appearance of localized states. There is however no distinct sharp peak in the simulated STS spectrum in the energy range from 0 to +1.0 eV. The propagation can be thus explained by the presence of a weakly localized $\pi$ states, where the localization is much weaker compared to the occupied region as shown by the isosurface plot in Fig. 3(e). Note that the feature of three-fold propagation in Fig. 3(e) is similar to the case of a single-atom vacancy in graphite [30, 41-44], though the localization is much weaker for the pyridinic-N.

**C. Graphitic-N**

The experimentally observed STM image at +500 mV and STS spectrum of a type B defect are shown in Fig. 4. The STS is different from that for the pristine graphite in terms of the asymmetric LDOS with respect to 0 V (the Fermi level), where larger intensity can be recognized in the positive bias region (unoccupied region). The small shoulder peak can be seen at the positive bias voltage, which is in contrast to that at the negative voltage observed for the pyridinic-N. The STS spectra with the same shape have been measured in the vicinity of the defect independent of the lateral position as measured in the pyridinic-N (see supplemental material, Figs. S9-S14). To identify the defect species in Fig. 4(a), DFT simulations of STM images and STS spectra were carried out assuming several types of defects structures. After comparison with the results of different defect models, it was found that the simulated STM and STS features of graphitic-N well reproduced those measured in experiments as shown in Figs. 4 and S3. That is, triangular bright spots in STM image and STS peak (enhanced intensity) at about +500 mV are common between experiment and theory. The defect was thus assigned as the graphitic-N. In the DFT calculation, the position of N atom in the graphitic-N species is almost the same as that of C atom in graphite (shorter by 0.002 nm for CN bond length than for CC bond length). That is, the nitrogen species takes the $sp^2$ planar structure of graphite. Concerning the origin of the STS peak at 500 mV, the state is ascribed to unoccupied π states ($p_z$ orbitals), which is similar to the edge state because of the localized character (see supplemental material, Fig. S3). In the simulated STM image (Fig. 5), the bright triangular spots are due to the enhanced LDOS component at about +500 mV. In addition, 15-20 carbon atoms are imaged bright, indicating that the localized π states propagate near the defect. Here, as shown in Figs. S4 and S5 in the supplemental material, the N atom position in Fig. 4(a) is found as the substituted position of a β-carbon atom (C atoms located above the centre of a hexagonal ring; the remaining C atoms are called α-carbon atoms [45]).

    Here, we discuss the appearance of the localized states in the unoccupied region. The positively charged N is expected by the chemical shift in N1s binding energy to higher energy for the graphitic-N (401.1 eV) in the XPS spectrum compared to that for the pyridinic-N (398.5 eV). Indeed, the difference of 2.6 eV is so large and the binding energy of 401.1 eV is

comparable with that for pyridinium ion, $C_5H_5NH^+$ (401.2 eV) and ammonium ion, $NH_4^+$ (401.5 eV), where N is positively charged in both cases [27], suggesting the positive charge of N of the graphitic-N. The positively charged N of the graphitic-N has also been reported recently based on the theoretical results by Yu et al. [8] and Mayer et al. [46]. The positive charge can be explained by electron transfer from N atom to the π conjugated state. If N is positively charged, the surrounding carbon atoms should be charged negatively because of the screening effect. The negative charge of carbon can explain a shift of the localized π state of carbon from the Fermi level to the upper energy level. In addition, the screening of positive N charge may be the source of the dark halo around the graphitic-N defect seen in Figs. 1(b) and (d), i.e. the charge density rearrangement may be induced by the positive charge of N and it may cause the modulated contrast in the STM image around N atom. Here we also note that the states with STS peak in the unoccupied region near the Fermi level may act as Lewis acid. The STS peak position of the graphitic-N should depend on the amount of doped nitrogen. The amount of electron doping ($n$) due to N impurities per unit area in a single graphene layer can be approximately quantified by the integral of the new (shifted) Fermi surface (*FS*) at a given energy (*E*) away from the pristine Fermi level. This, for independent electrons, is represented as

$$n = \frac{4}{4\pi^2} \oint_{FS} d\mathbf{k} = \frac{|k|^2}{\pi} = \frac{E^2}{\hbar^2 \pi v_F^2} \qquad (1)$$

in which we included the factor 4 to account for spin and valley degeneracy and where the linear band dispersion of graphene $E = \hbar v_F |k|$ was used. Then, the shift in the Fermi energy should scale as $E = \hbar v_F (\pi n)^{1/2}$, where $v_F$ is the Fermi velocity (about $10^6$ m s$^{-1}$). Following these arguments, it may be possible to control the population of the defect states by controlling the N concentration. For low N doses (very small *n*), the extra electron introduced by the impurity into the graphene layer would cause infinitely small doping, thus inducing a positive charge on the N atom. Otherwise, for high doses, the doping would be sufficient to doubly occupy the $p_z$ orbital (i.e. the states responsible for the STS peak at +500 mV), giving rise to a neutral defect. From Eq. (1), we estimated that a concentration of more than $1.8 \times 10^{13}$ defects cm$^{-2}$ (about a graphitic-N every 210 carbon atoms in the single graphene layer) should be sufficient to raise the Fermi energy by about 0.5 eV, where these states occur.

**V. Summary and conclusions**

We have examined a nitrogen-doped graphite surface by scanning tunnelling microscopy, scanning tunnelling spectroscopy, X-ray photoelectron spectroscopy, and first-principles calculations. Two types of nitrogen species have been identified on the nitrogen-doped graphite surface: pyridinic-N with single-atom vacancy and graphitic-N. While pyridinic-N defects induce an atomic rearrangement to form a pentagon, graphitic-N affects the structure only slightly. In each case, the electronic structure of graphite close to the Fermi energy is found to be modified by the defects. The distinct localized π states appear in occupied and unoccupied region near the Fermi level around pyridinic-N and graphitic-N species, respectively.

This work was financially supported by the New Energy and Industrial Technology Development Organization of Japan. We acknowledge CINECA consortium for the computational resources and Dr. Rocco Martinazzo for fruitful discussions.

*Figure captions*

FIG. 1 (colour online)

(a) N1s XPS spectra of graphite surface after nitrogen ion bombardment at 200 eV. The results before and after annealing at 900 ± 50 K for 300 s are shown. Deconvoluted components are pyridinic-N (398.5 eV; before: 32.6%, after: 31.4%), pyrrolic-N (399.9eV; before: 17.4%, after: 7.6%), graphitic-N (401.1eV; before: 40.1%, after: 54.3%), and oxide-N (403.2eV; before: 9.8%, after: 6.8%). (b) Typical STM topographic image of the graphite surface at about 5.3 K (scan size: 49.45 × 46.97 nm$^2$, tunnelling current $I_t$ = 179 pA, sample bias $V_s$ = 98.8 mV). (c)(d) STM topographic images of regions A and B in Fig. 1(b), respectively (in both case, scan size: 9.88 × 9.26 nm$^2$, $I_t$ = 97.5 pA, $V_s$ = -109 mV).

FIG. 2 (colour online)

(a) STM topographic image of region A in Fig. 1(b) (scan size: 4.81 × 4.61 nm$^2$, $I_t$ = 96.9 pA, $V_s$ = -108 mV). The simulated STM image ($V$ = -0.1 V) is also shown for comparison. (b) STS spectrum measured at the position indicated by the arrow in (a). (c) The equilibrium geometry of pyridinic-N defect calculated by DFT (d) Simulated STS spectrum of pyridinic-N.

FIG. 3 (colour online)

(a)(d) Simulated STM images (constant-current mode) for pyridinic-N graphene defect. (a) $V$ = -1.0 V, (d): $V$ = +1.0 V. The N atom is placed at the centre of the image. (b) DFT equilibrium geometry and isosurface plot of electron density at $1.5 \times 10^{-3}$ electrons/Å$^3$ integrated from -0.7 eV to Fermi level (square modulus of wave function up to Fermi level from -0.7 eV). The N atom is shown in light grey colour. (c) Same image as Fig. 3(b) except for colour near nitrogen atom, where only *xy*-plane DOS contribution near N is represented by blue-red colours. (e) DFT equilibrium geometry and isosurface plot of electron density at $1.5 \times 10^{-3}$ electrons/Å$^3$ integrated from Fermi level to +0.7 eV. (f) Schematic representation of p$_z$ orbitals in pyridinic-N.

FIG. 4 (colour online)

(a) STM topographic image corresponding to the defect shown in region B in Fig. 1(b) (image was taken from different sample, scan size: 5.09 × 5.08 nm$^2$, $I_t$ = 39.0 pA, $V_s$ = 500 mV). The simulated STM image (V = +0.5 V) is also shown for comparison. (b) STS spectrum measured at the position indicated by the arrow in (a). (c) The equilibrium geometry of Graphitic-N defect calculated by DFT (d) Simulated STS spectrum of graphitic-N.

FIG. 5 (colour online)

(a)(c) Simulated STM images of graphitic-N defect. (a) V = -0.5 V, (c) V = +0.5 V. The N atom is placed at the centre of the image. (b)(d) DFT equilibrium geometry and isosurface plot of electron density, (b) electron density at $2\times10^{-4}$ electrons/Å$^3$ integrated from -0.7 eV to Fermi level (square modulus of wave function up to Fermi level from -0.7 eV), (d) electron density at $1.5\times10^{-3}$ electrons/Å$^3$ integrated from Fermi level to +0.7 eV.

Figure 1

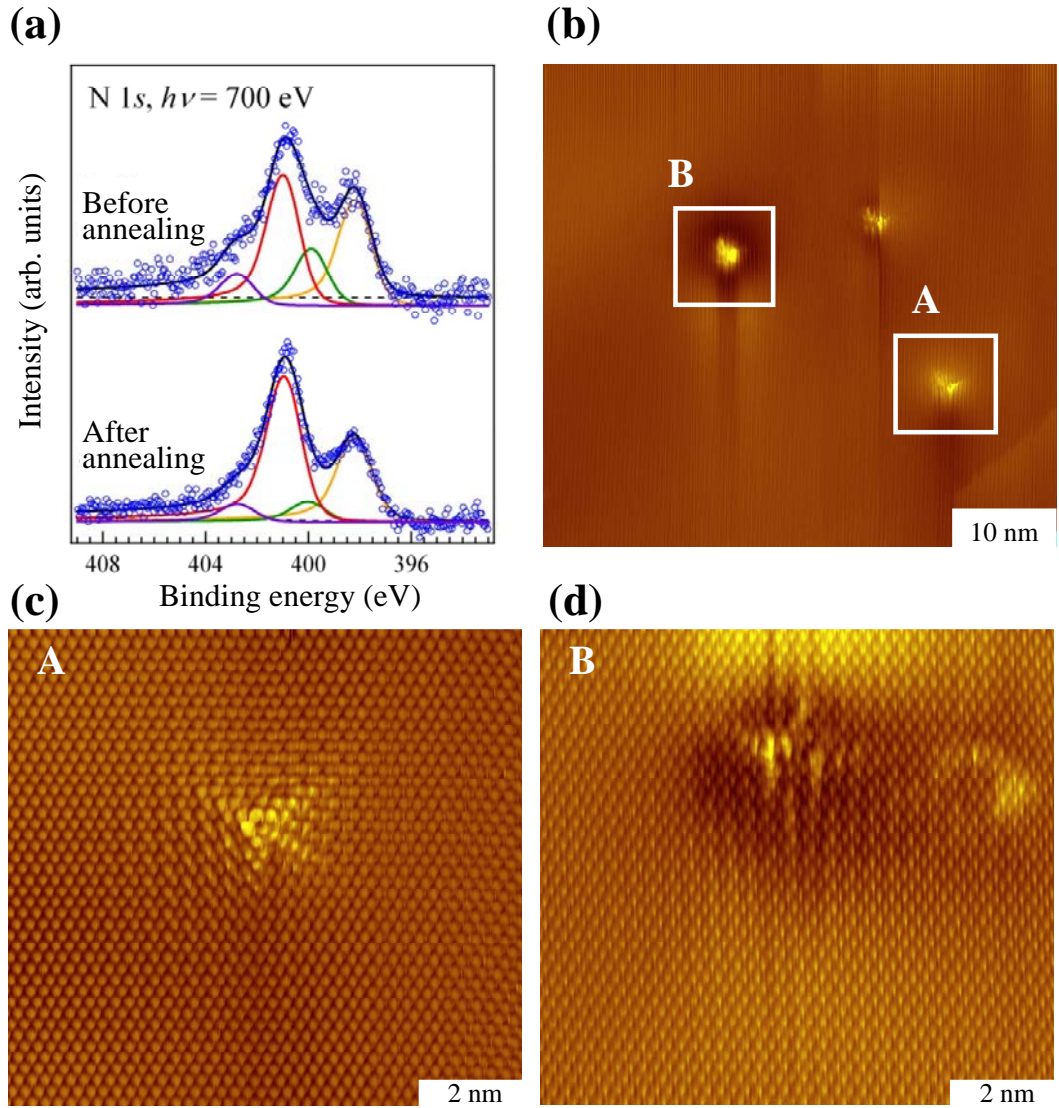

FIG. 1 (colour online)

(a) N1s XPS spectra of N-doped graphite surface after nitrogen ion bombardment at 200 eV. The results before and after annealing at 900 ± 50 K for 300 s are shown. Deconvoluted components are pyridinic-N (398.5 eV; before: 32.6%, after: 31.4%), pyrrolic-N (399.9eV; before: 17.4%, after: 7.6%), graphitic-N (401.1eV; before: 40.1%, after: 54.3%), and oxide-N (403.2eV; before: 9.8%, after: 6.8%). (b) Typical STM topographic image of the graphite surface at about 5.3 K (scan size: 49.45 × 46.97 nm$^2$, tunnelling current $I_t$ = 179 pA, sample bias $V_s$ = 98.8 mV). (c)(d) STM topographic images of regions A and B in Fig. 1(b), respectively (in both case, scan size: 9.88 × 9.26 nm$^2$, $I_t$ = 97.5 pA, $V_s$ = -109 mV).

Figure 2

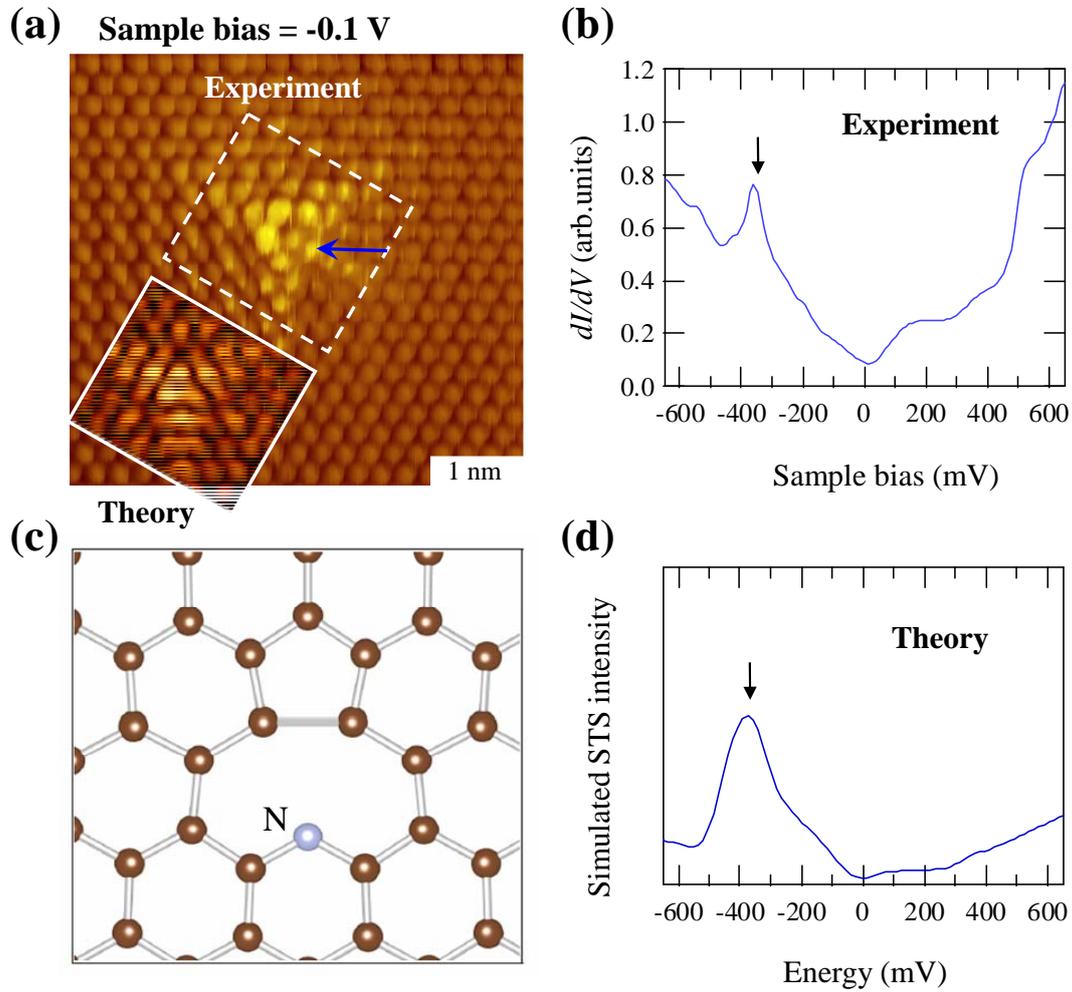

FIG. 2 (colour online)

(a) STM topographic image of region A in Fig. 1(b) (scan size: 4.81 × 4.61 nm$^2$, $I_t$ = 96.9 pA, $V_s$ = -108 mV). The simulated STM image ($V$ = -0.1 V) is also shown for comparison. (b) STS spectrum measured at the position indicated by the arrow in (a). (c) The equilibrium geometry of pyridinic-N defect calculated by DFT (d) Simulated STS spectrum of pyridinic-N.

Figure 3

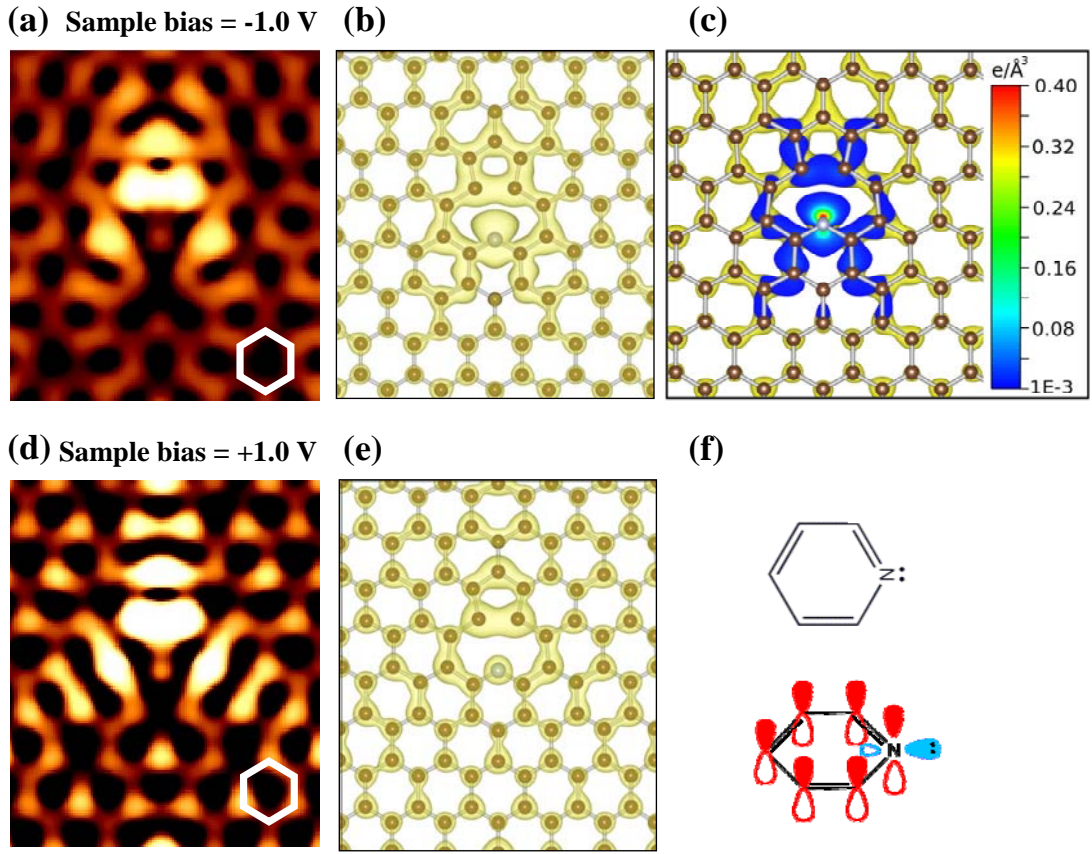

FIG. 3 (colour online)

(a)(d) Simulated STM images (constant-current mode) for pyridinic-N graphene defect. (a) $V = -1.0$ V, (d): $V = +1.0$ V. The N atom is placed at the centre of the image. (b) DFT equilibrium geometry and isosurface plot of electron density at $1.5 \times 10^{-3}$ electrons/Å$^3$ integrated from $-0.7$ eV to Fermi level (square modulus of wave function up to Fermi level from $-0.7$ eV). The N atom is shown in light grey colour. (c) Same image as Fig. 3(b) except for colour near nitrogen atom, where only *xy*-plane DOS contribution near N is represented by blue-red colours. (e) DFT equilibrium geometry and isosurface plot of electron density at $1.5 \times 10^{-3}$ electrons/Å$^3$ integrated from Fermi level to $+0.7$ eV. (f) Schematic representation of p$_z$ orbitals in pyridinic-N.

Figure 4

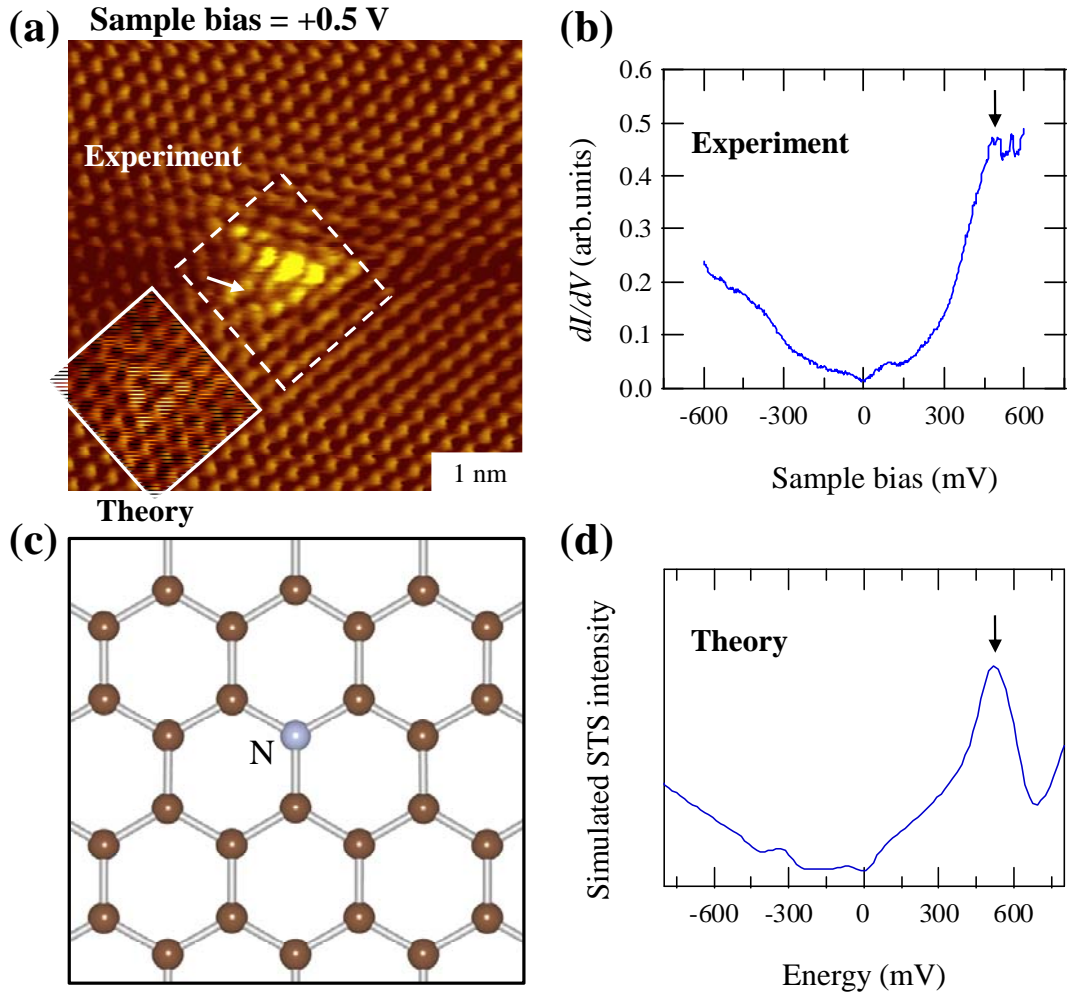

FIG. 4 (colour online)

(a) STM topographic image corresponding to the defect shown in region B in Fig. 1(b) (image was taken from different sample, scan size: 5.09 × 5.08 nm$^2$, $I_t$ = 39.0 pA, $V_s$ = 500 mV). The simulated STM image (V = +0.5 V) is also shown for comparison. (b) STS spectrum measured at the position indicated by the arrow in (a). (c) The equilibrium geometry of Graphitic-N defect calculated by DFT (d) Simulated STS spectrum of graphitic-N.

Figure 5

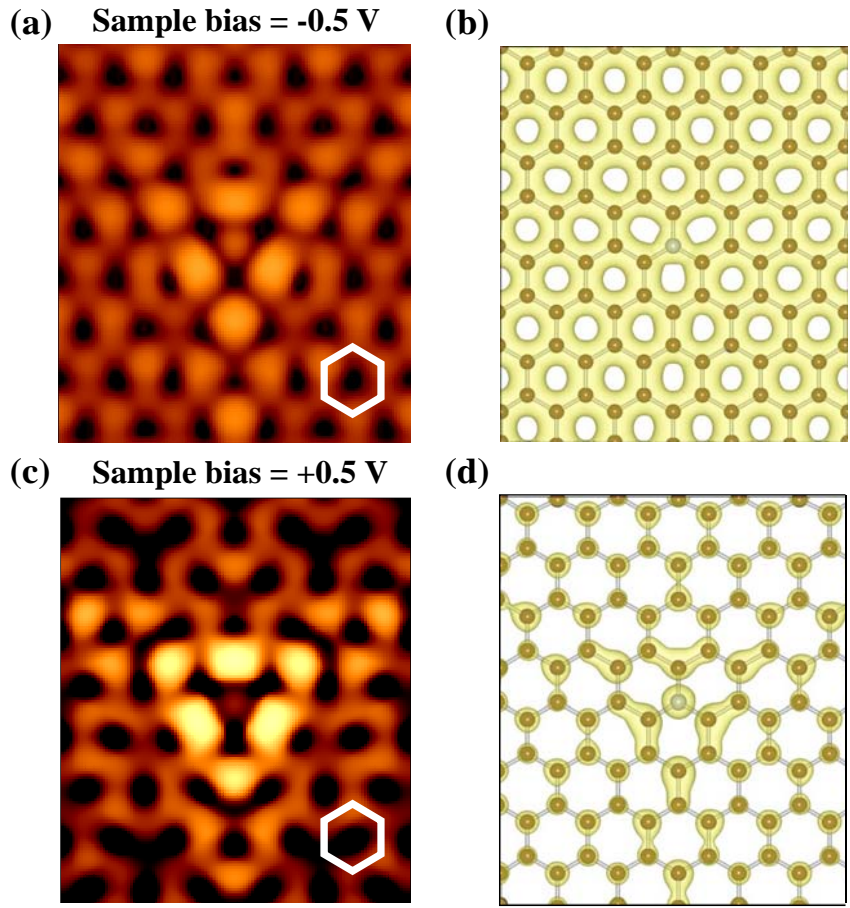

FIG. 5 (colour online)

(a)(c) Simulated STM images of graphitic-N defect. (a) V = -0.5 V, (c) V = +0.5 V. The N atom is placed at the centre of the image. (b)(d) DFT equilibrium geometry and isosurface plot of electron density, (b) electron density at $2\times 10^{-4}$ electrons/Å$^3$ integrated from -0.7 eV to Fermi level (square modulus of wave function up to Fermi level from -0.7 eV), (d) electron density at $1.5\times 10^{-3}$ electrons/Å$^3$ integrated from Fermi level to +0.7 eV.

# Supplemental material

# Atomic-scale characterization of nitrogen-doped graphite: Effects of the dopant nitrogen on the local electronic structure of the surrounding carbon atoms

Takahiro Kondo[1]†, Simone Casolo[2]†, Tetsuya Suzuki[1], Taishi Shikano[1], Masataka Sakurai[1], Yoshihisa Harada[3,4], Makoto Saito[3,††], Masaharu Oshima[3,4], Mario Italo Trioni[5], Gian Franco Tantardini[2,5] and Junji Nakamura[1]∗

† The authors contributed equally

∗ corresponding author: nakamura@ims.tsukuba.ac.jp

1 Faculty of Pure and Applied Sciences, University of Tsukuba, 1-1-1 Tennodai, Tsukuba, Ibaraki 305-8573, Japan

2 Dipartimento di Chimica Fisica ed Elettrochimica, Università degli Studi di Milano, via Golgi 19, 20133 Milan, Italy

3 Department of Applied Chemistry, The University of Tokyo, 7-3-1 Bunkyo-ku, Tokyo 113-8656, Japan

4 The University of Tokyo Synchrotron Radiation Research Organization, 7-3-1 Bunkyo-ku, Tokyo 113-8656, Japan

5 CNR National Research Council of Italy, ISTM, via Golgi 19, 20133 Milan, Italy

## Contents

In this supplemental material, at first, we describe the efficiency of N-doping on graphite by nitrogen ion bombardment. Then, we describe the effect of the bias on the STM image and results of analysis of the exact position of N atom ($\alpha$- or $\beta$-carbon positions) in N-doped graphite. Finally, we demonstrate an example of the characterization of N species on N-doped graphite based on this work.

††present address: Toyota Motor. Corporation, 1200, Mishuku, Susono, Shizuoka, 410-11 Japan

# 1. Efficiency of N-doping on graphite by ion bombardment

We have examined the efficiency of nitrogen-doping by the ion bombardment method by taking STM images of N-doped graphite. The N-doped graphite used here was prepared by nitrogen ion bombardment with 200 eV ions at normal incidence at 300 K, followed by 10 min annealing at $900\pm50$ K. Total irradiated ion amount per unit area $N$ was estimated as $0.019\pm0.010$ nm$^{-2}$ by using Equation (1) with the ion current $I = 2\pm1$ nA, which is independently measured by the Faraday cup.

$$N = \frac{I \cdot t}{e \cdot A_F} \quad (1)$$

Here, $t$ is the total time of the bombardment (30 s), $e$ is elementary charge, and $A_f$ is the area of the Faraday cup (19.6 mm$^2$).

Fig. S1 shows the series of STM images (larger and smaller scan area) taken at different sample positions, where every position was separated by more than 1 μm. By counting bright spots in image, we have estimated the defect density of 0.016, 0.014, 0.017 and 0.016 for Figs. 1(a,e), 1(b,f), 1(c,g) and (d,h), respectively. Those are about 70-90 % of the amount of the total irradiated ions, indicating high efficiency of defects formation.

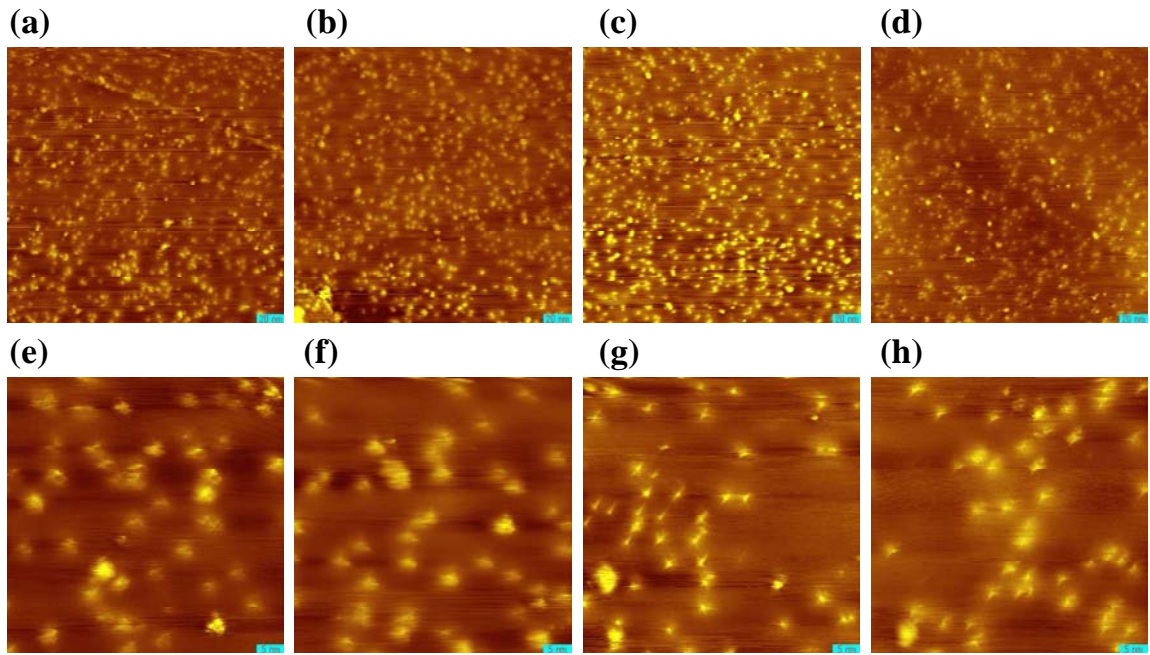

**FIG. S1** STM images of N-doped graphite at 67 K, **(a)-(d)** 200 × 200 nm$^2$, $I_t$ = 180 pA, $V_s$ = 300, **(e)-(h)** 50 × 50 nm$^2$, $I_t$ = 60 pA, $V_s$ = 300 mV

## 2. Effect of the sample bias on the STM images
### (1) Pyridinic-N defect in graphite

Simulated STM images with different bias condition of pyridinic-N defect in graphite are shown in Fig. S2(a), together with the experimental STM image (Fig. S2(b)). At the lowest bias of -0.1 eV, simulated STM image shows a bright region propagating toward three directions due to the localized π states. When the bias becomes larger (more negative) than -0.50 eV, this propagating feature becomes weaker and a two-fold bright feature located at the pentagon C atoms and their neighbors is left, as shown in the images taken at -0.75 eV and -1.0 eV in Fig. S2(a). In these conditions, the bright feature is dominated by the contribution by the DOS forming a large STS peak at -370 mV. Based on the DFT, the states can be assigned as the contributions of $p_z$ orbitals of N atom and C atoms especially forming the pentagon. Fig. S2(c) shows the $p_z$ contribution of DOS (projected LDOS of $p_z$ contribution). Projected LDOS for N, C-meta (Carbon atom at meta position with respect to N) and one of two pentagon C atoms have a large peak component at around -370 meV, suggesting that these states are the origin of the observed STS peak at -370 mV.

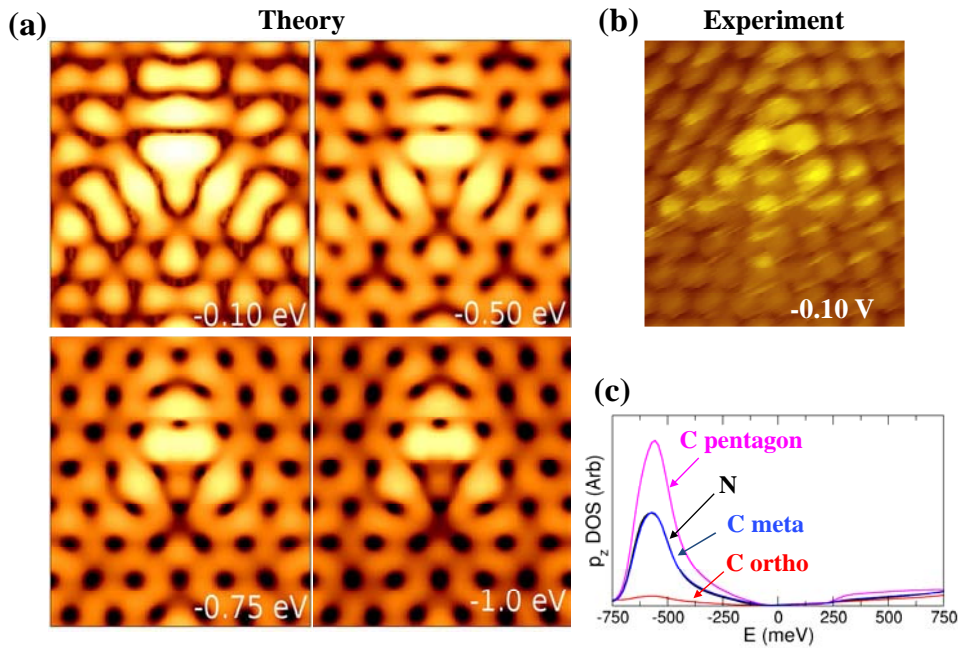

**FIG. S2 (a)** Simulated STM images at different bias conditions, **(b)** Experimental STM image at the sample bias of -0.1 V (same as Fig. 1(a)), **(c)** Projected LDOS of $p_z$ contribution for the N atom C pentagon (one of two C atoms forming the pentagon bond), C-meta (Carbon atom at meta position with respect to N) and C-ortho (Carbon atom at nearest-neighbors to N) in pyridinic-N.

### (2) Graphitic-N defect in graphite

Experimentally observed STM images of graphitic-N defect in graphite are shown in Fig. S3, together with simulated STM images and STS spectrum. Contrary to the case of Fig. 1, there is no dark-halo pattern around the nitrogen at the sample bias of +500 mV as shown in Figs. S3(a) and 3(b). On the other hand, a dark-halo was still observed at the sample bias of -500 mV (Fig. S3(d)) as in the cases of +98.8 mV and -109 mV in Fig. 1. As described in section IVC, the source of the dark halo around the graphitic-N is considered to be caused by the screening of the Coulomb potential generated by the positively charged N.

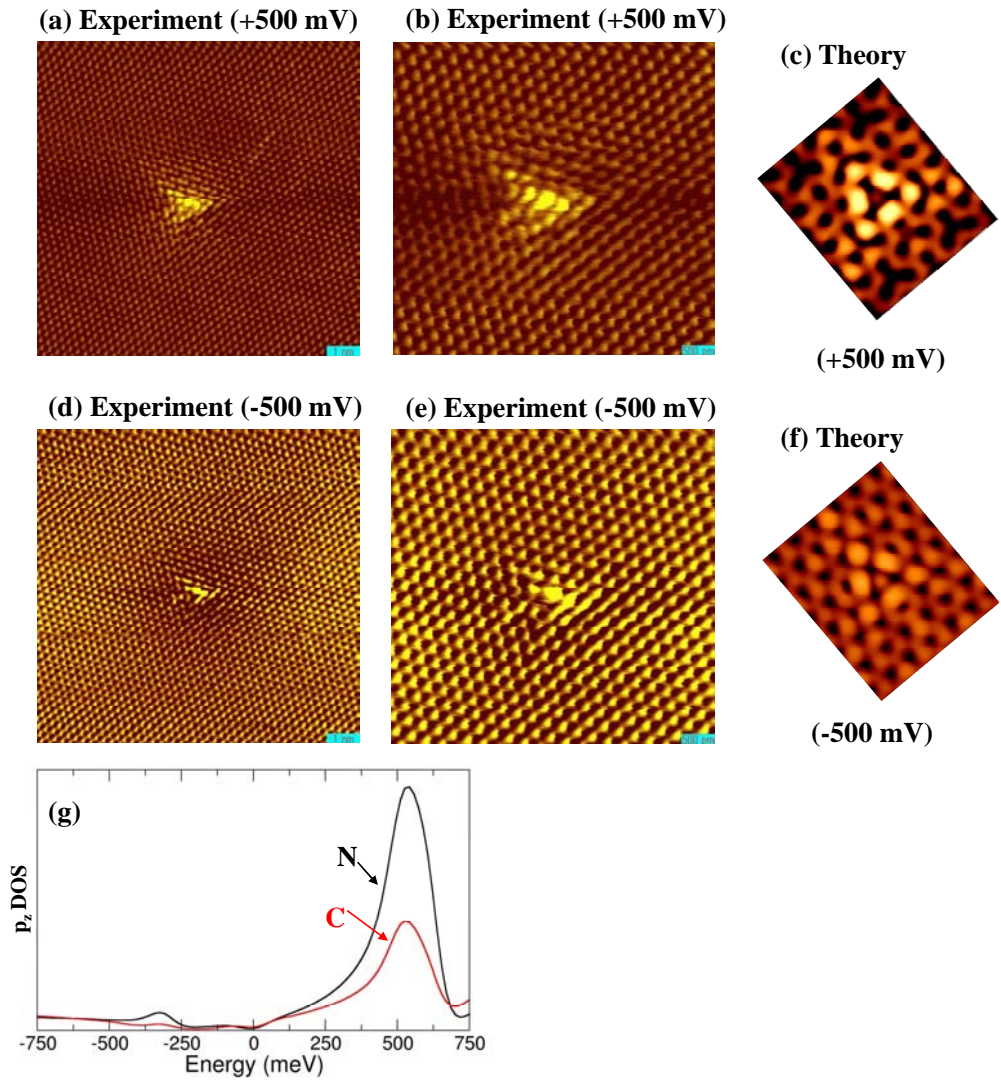

**FIG. S3** STM images around graphitic-N defect in graphite: **(a)** $10.5 \times 10.2$ nm$^2$, +500 mV, 40 pA, **(b)** $5.25 \times 5.1$ nm$^2$, +500 mV, 40 pA, **(c)** Simulated STM image (+500 mV) **(d)** $10.5 \times 10.2$ nm$^2$, -500 mV, 40 pA **(e)** $5.25 \times 5.1$ nm$^2$, -500 mV, 40 pA, **(f)** Simulated STM image (-500 mV). **(g)** Projected LDOS of p$_z$ contribution for the N atom and that for one of three nearest neighbors C atoms with respect to N in graphitic-N.

The bright spots in Fig. S3(c) are dominated by the contribution of DOS forming a large STS peak at +500 mV. Based on the DFT, the states can be assigned as the contributions of $p_z$ orbitals of N atom and C atoms. Fig. S3(g) shows the $p_z$ contribution of DOS (projected LDOS) at N and C (one of the nearest neighbors C atoms with respect to N). Both have a large peak component at around +500 meV, suggesting that these states are the origin of the observed STS peak at +500 mV.

## 3. Analysis and effect of N atom position ($\alpha$- or $\beta$-carbon positions)
### (1) Analysis of N position ($\alpha$- and $\beta$-carbon positions)

The STM image of graphitic-N is shown in Fig. S4 together with the calculated STM image, where a hexagonal mesh has been superimposed on the image in order to identify $\alpha$- and $\beta$-carbon positions. $\beta$-carbon atoms are marked by white-balls on the mesh ($\beta$-carbon: the carbon atoms located above the centers of the hexagons of the graphene layer beneath; the remaining carbon atoms are $\alpha$-carbon atoms.) It is known that the $\beta$ positions appear brighter in STM within the 6 member rings of graphite [1]. To analyze the position of N, therefore, we have put the mesh on the STM image in coincidence with bright spots far away from the defect in the STM image. Here, we note that the bright spots near the center of defect do not coincide with the white-balls of the mesh because the electronic structure near the center of the defect is modified by the defect.

Experimental STM image is well reproduced by the calculations in which N substitutes a $\beta$-carbon position, here shown by a green ball, suggesting that Fig. 3a is the STM image of a graphitic-N defect in $\beta$-position.

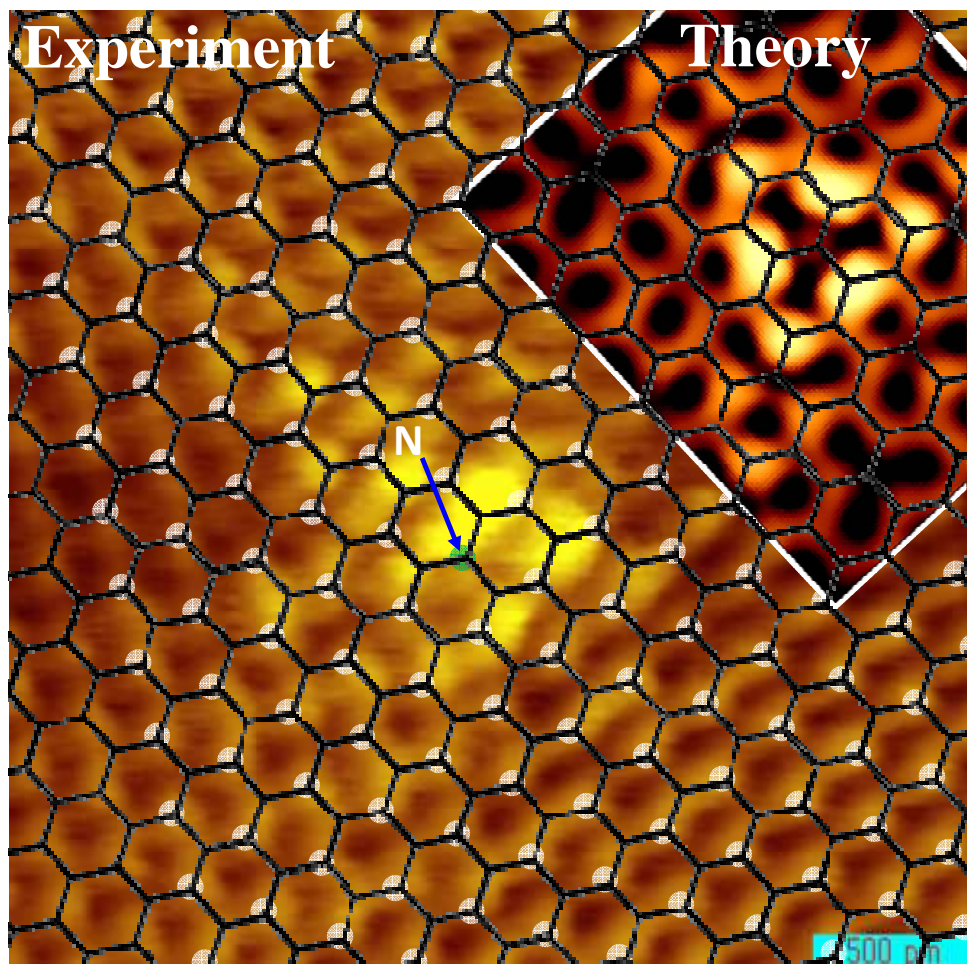

**FIG. S4** Comparison of STM images between experiment and theory for graphitic-N. To identify the position of N in the graphitic-N, the black hexagonal mesh is superimposed on the image, where white-balls represent the $\beta$-carbon positions. The analysis suggests this STM is corresponds to the graphitic-N defect in $\beta$-position.

### (2) Effect of N position ($\alpha$- and $\beta$-carbon positions) on STS spectrum

It can be expected that DOS are different between for $\alpha$-position N and for $\beta$-position N defects, because the $p_z$ orbitals of N and C in the beneath layer have an overlap in the case of $\alpha$-position N, while in the case of $\beta$-position, they don't. To examine the difference in DOS between $\alpha$-position N and $\beta$-position N, the theoretical STS spectra for both the pyridinic-N graphitic-N defects were calculated and are shown in Fig. S5. In the case of pyridinic-N, the main peak locates 70 mV deeper for $\alpha$-position possibly due to the effect of the overlap of $p_z$ orbitals between N and C of the beneath layer. On the other hand, for the graphitic-N, the $\beta$-position spectrum shows a 15 mV higher shift for the main STS peak at positive bias. Overall, the differences between $\alpha$ and $\beta$ positions are small and we expect them to be difficult to be distinguished in STS experiments. Instead, the doping at $\alpha$/$\beta$ positions can be characterized by identifying the two sublattices from STM images, as shown in Fig. S4.

Our simulations predict that pyridinic-N defect in the $\beta$-position N (hence with the vacancy in $\alpha$) is 15 meV more stable than that in the $\alpha$-position. Also, it is found that the graphitic-N defect in $\beta$-position is 25 meV more stable than that in $\alpha$-position at the nitrogen concentrations used in this study. In order to experimentally clarify the stable N-position, we have counted each defect by STM. However, we could not collect a sufficient amount of data to evaluate the $\alpha$ and $\beta$ populations; we have identified 27 times for graphitic-N and 6 times for pyridinic-N by STM/STS, but these are not always measured with atomic resolution. This is mainly due to the difficulty of the STM/STS measurement in our system due to the apparatus conditions such as the STM-tip condition. We could identify the $\alpha$ and $\beta$ positions only 2 times for graphitic-N (one is $\alpha$ and the other is $\beta$).

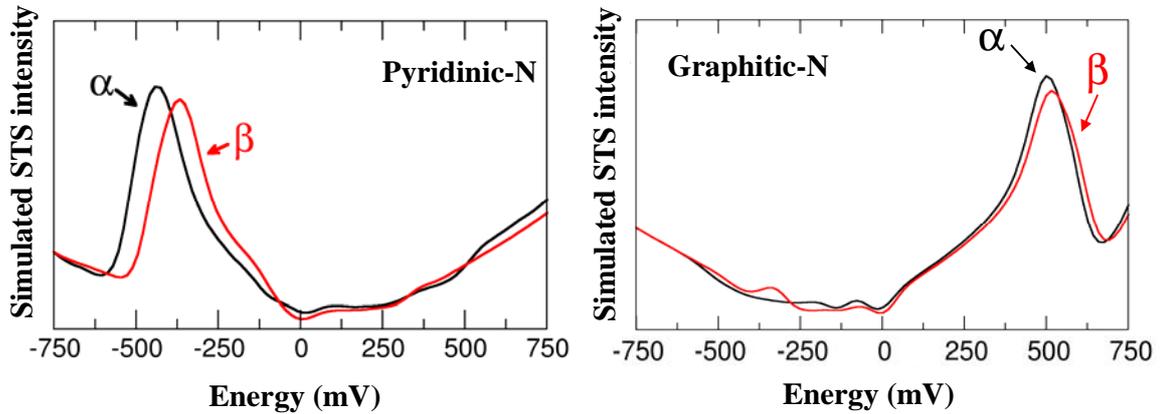

**FIG. S5** Theoretical STS simulation for the $\alpha$-position N and $\beta$-position N of graphitic-N.

**(3) Small peaks in the STS spectrum**

Two small features at around the Fermi energy (-100 and +100 mV) in the STS spectrum in Figs. 2(b), 2(d), 4(b), 4(d) and S5 can be assigned as peaks originated from the interaction of the defective layer with the underneath layer, because these peaks are not observed in the simulated STS spectrum of the same defect in a single graphene layer, while peaks appeared in the cases of both the α and β defect positions in graphite as shown in Fig.S5.

## 4. STS around defect of nitrogen doped graphite

STS spectra taken around the pyridinic-N defect of N-doped graphite are shown in Figs. S6 and S7. Peak intensity and peak energy in STS around -370 mV are plotted as a function of the position in Fig. S8. The same set for the graphitic-N is shown in Figs. S9- S14. In both pyridinic-N and graphitic-N cases, there are no significant spatial dependences of both intensity and energy. Only slight oscillations of the peak intensity and energy position as a function of the distance are observed. This is possibly due to the interaction between the localized $p_z$ orbital (peak component in STS) and the π-band electrons of graphite [2]. Small position-dependence indicates no significant anisotropic perturbation of the electronic structure of the surface, which is different from the case of the single vacancy defect of carbon on graphite, where a large anisotropic perturbation is observed [2].

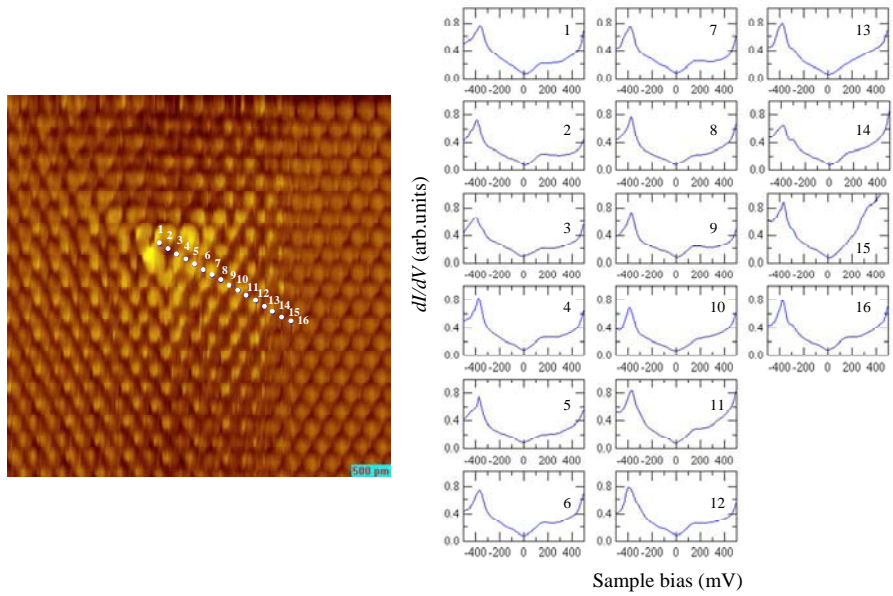

**FIG. S6** STS spectra around pyridinic-N of N-doped graphite in Fig. 1(c) STS spectra taken at the position labeled in the left STM image are shown at right. (Sample bias -108 mV, current 97.9 pA)

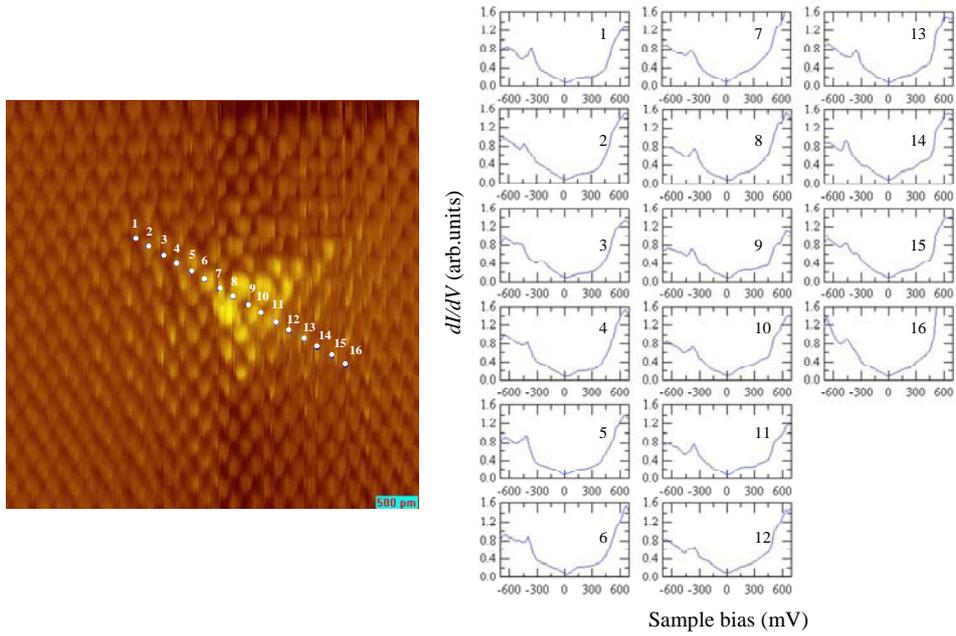

**FIG. S7** STS spectra around pyridinic-N of N-doped graphite in Fig. 1(c) STS spectra taken at the position labeled in the left STM image are shown at right. (Sample bias -109 mV, current 96.2 pA)

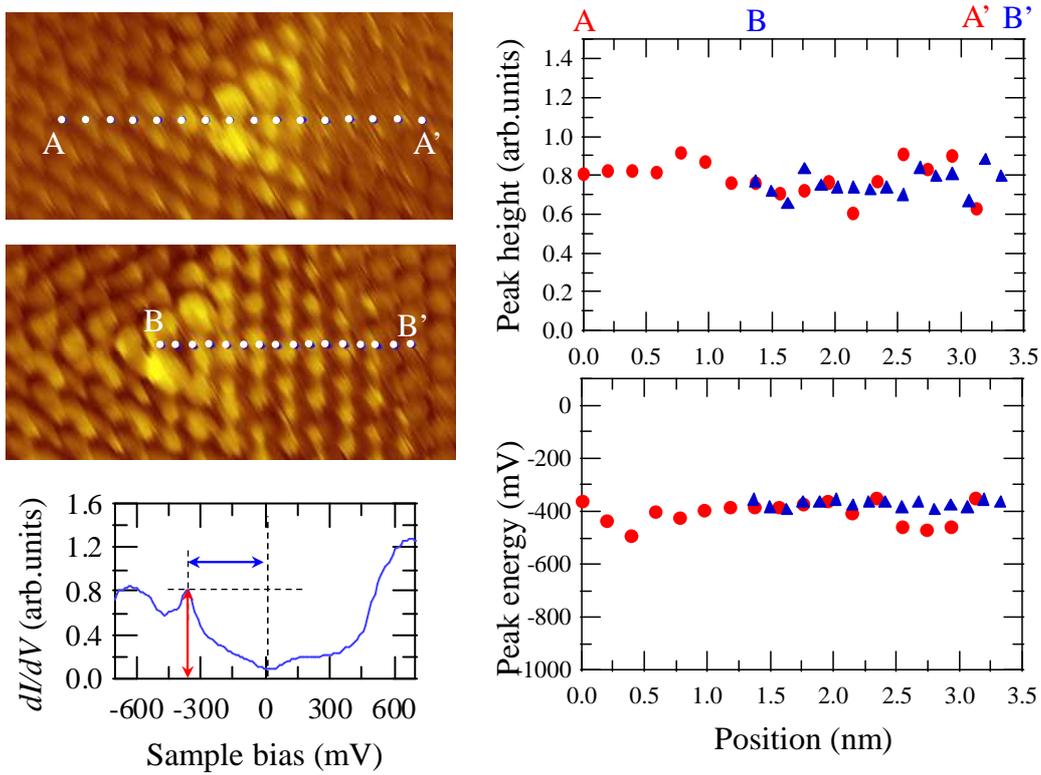

**FIG. S8** STS spectra around pyridinic-N of N-doped graphite in Fig. 1(c) STS peak intensity and peak energy are plotted as a function of distance. Red filled circles represent the peak height and energy at the position indicated in the region between A and A'. Blue triangles represent those in the region between B and B'

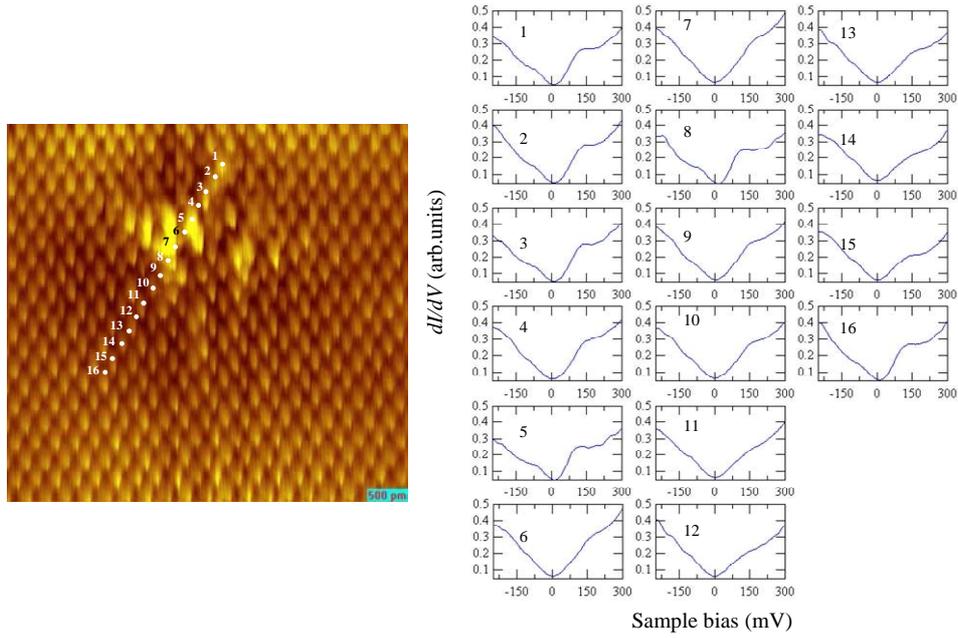

**FIG. S9** STS spectra around graphitic-N of N-doped graphite in Fig. 1(d) STS spectra taken at the position labeled in the left STM image are shown at right. (STS spectra were taken with the tip-sample distance determined by the condition with the sample bias of -108 mV and current 96.5 pA, STM image was taken with the sample bias of -108 mV and current 97.4 pA )

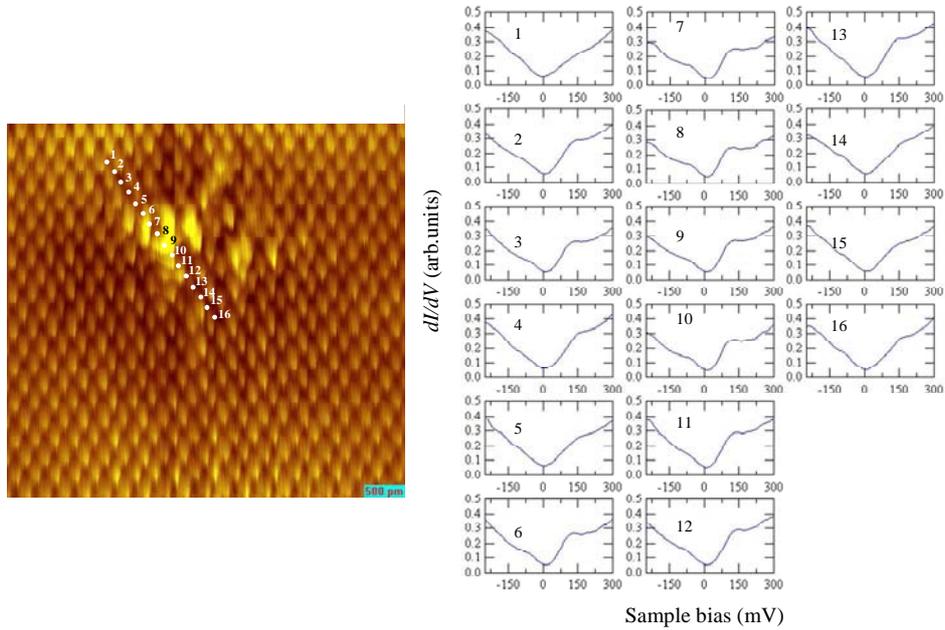

**FIG. S10** STS spectra around graphitic-N of N-doped graphite in Fig. 1(d) STS spectra taken at the position labeled in the left STM image are shown at right. (STS spectra were taken with the tip-sample distance determined by the condition with the sample bias of -108 mV and current 96.3 pA, STM image was taken with the sample bias of -108 mV and current 97.4 pA )

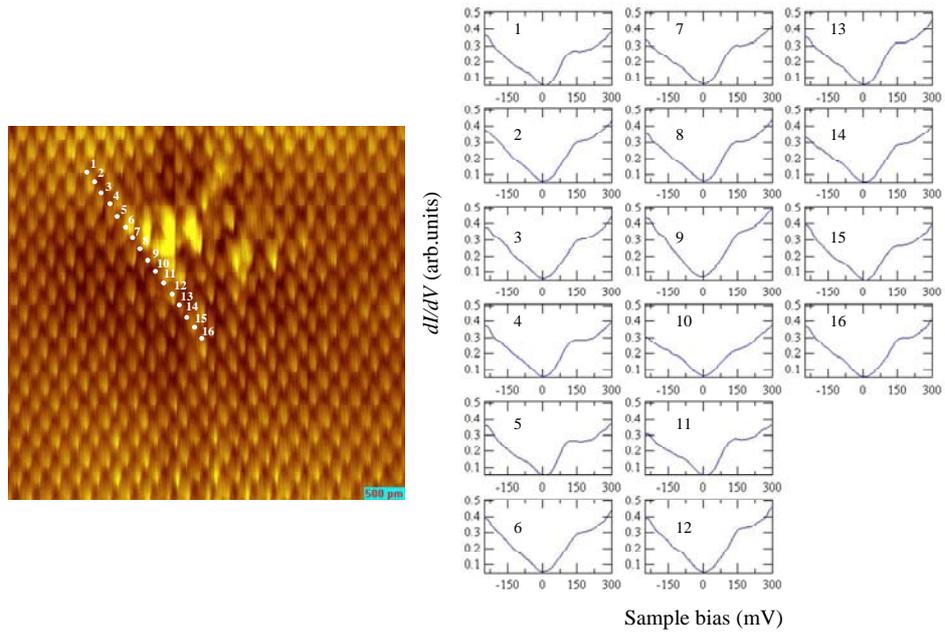

**FIG. S11** STS spectra around graphitic-N of N-doped graphite in Fig. 1(d) STS spectra taken at the position labeled in the left STM image are shown at right. (STS spectra were taken with the tip-sample distance determined by the condition with the sample bias of -108 mV and current 97.0 pA, STM image was taken with the sample bias of -108 mV and current 97.4 pA )

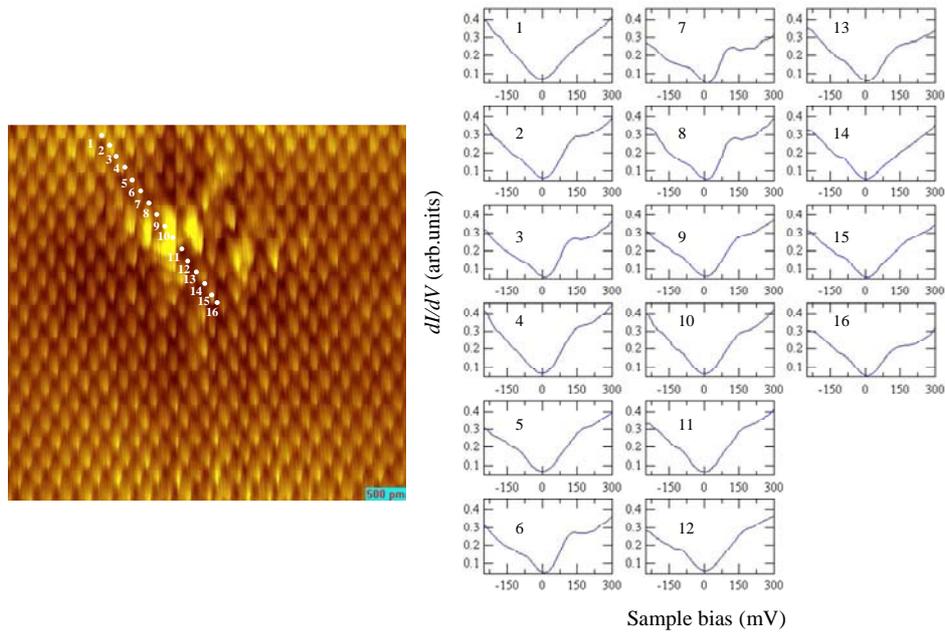

**FIG. S12** STS spectra around graphitic-N of N-doped graphite in Fig. 1(d) STS spectra taken at the position labeled in the left STM image are shown at right. (STS spectra were taken with the tip-sample distance determined by the condition with the sample bias of -108 mV and current 97.3 pA, STM image was taken with the sample bias of -108 mV and current 97.4 pA )

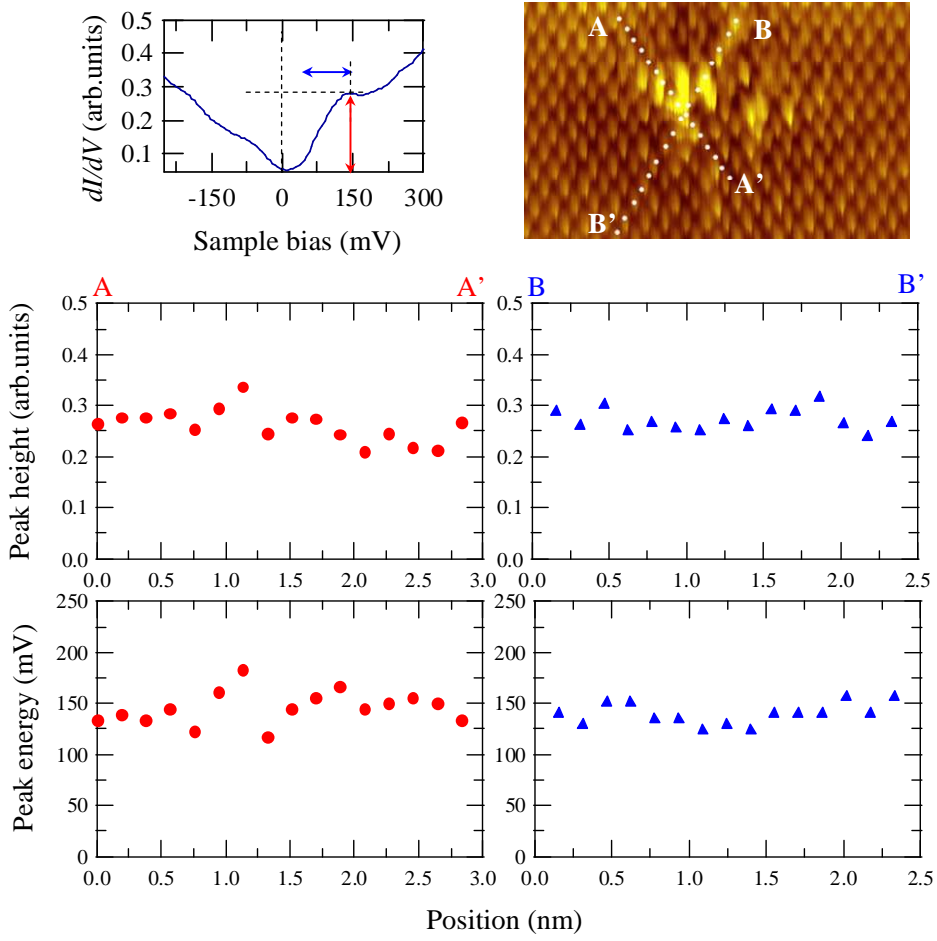

**FIG. S13** STS spectra around graphitic-N of N-doped graphite in Fig. 1(d) STS peak intensity and peak energy are plotted as a function of distance Red filled circles represent the peak height and energy at the position indicated in the region between A and A'. Blue triangles represent those in the region between B and B'

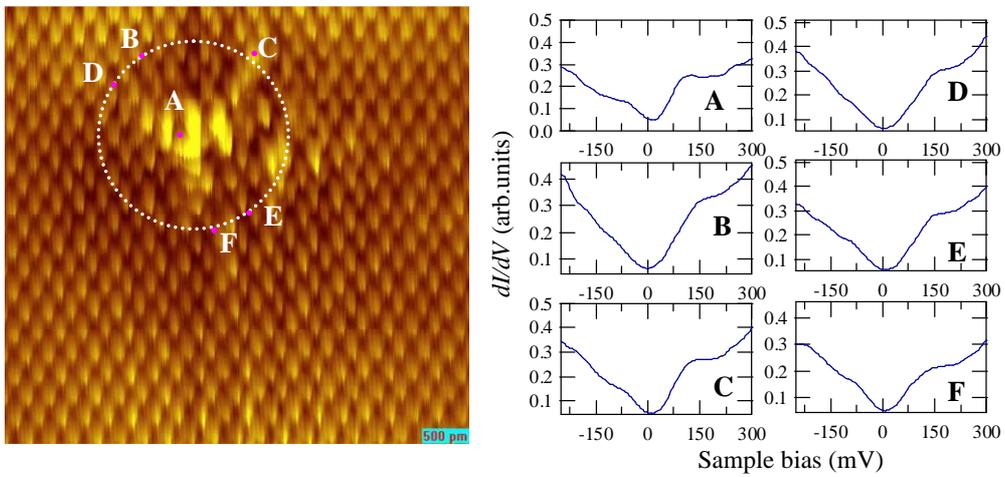

**FIG. S14** STS spectra around graphitic-N of N-doped graphite in Fig. 1(d)

**5. Example of the characterization of N species on N-doped graphite**

Fig. S15 shows the STM image and STS spectrum of other types of defective structures on N-doped graphite. The STM image was taken at -100 mV bias. In this particular case, two types of defects co-exist within a few nanometers of each other. Defect-A in Fig. S15 has a bright region in STM image that propagates in three directions. This feature is similar to that of the single-atom vacancy defect in graphite [2] and that of the pyridinic N with vacancy in Fig. 2. Because the only single distinct STS peak appears at the unoccupied state for defect-A in Fig. S15, it can be assigned to a single-atom vacancy defect. On the other hand, defect-B in Fig. S15 has a shape very similar to the graphitic N defect shown in Fig. 4, although a small peak at +60 mV was found in this case. Therefore, we could not clearly identify this defect, but we suggest the possibility that B is a graphitic N defect whose STM images and STS spectra were possibly modified by the presence of the nearby vacancy.

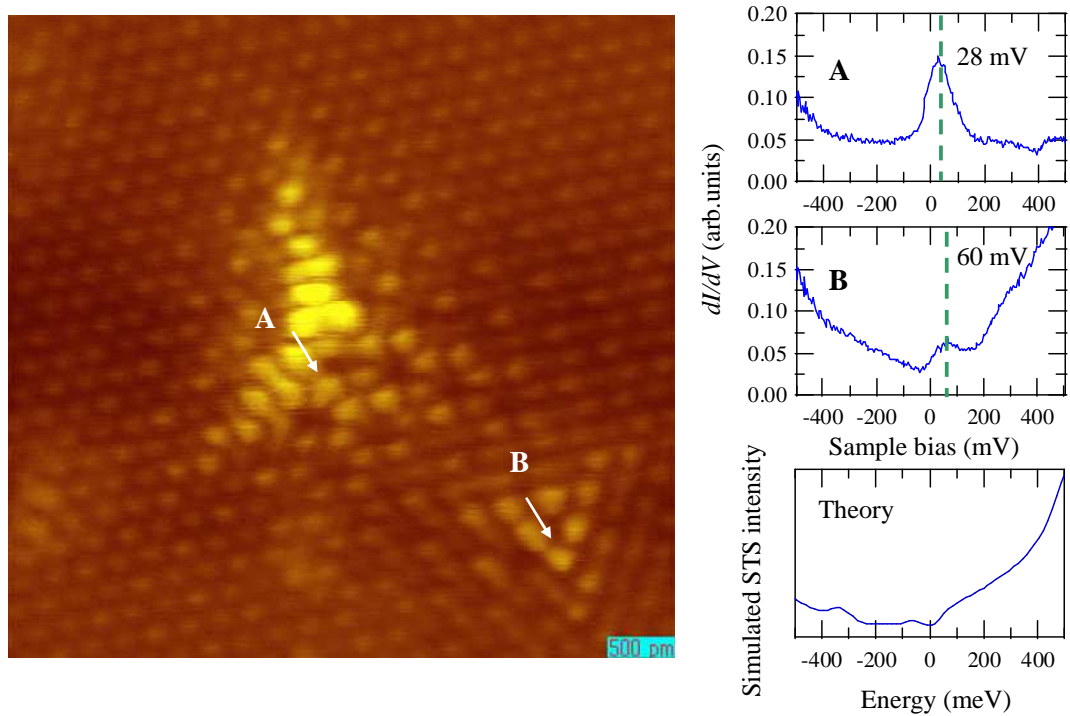

**FIG. S15 (Left)** STM topographic image (scan size: 5.25 × 5.1 nm$^2$, $I_t$ = 38.6 pA, $V_s$ = -500 mV). **(Right) A and B**: STS spectra measured at positions A and B in the STM image, respectively. The simulated STS spectrum (same as Fig. 4(d)) is also shown. (STS spectra were taken with the tip-sample distance determined by the condition with the sample bias of -499 mV and current 38.9 pA)

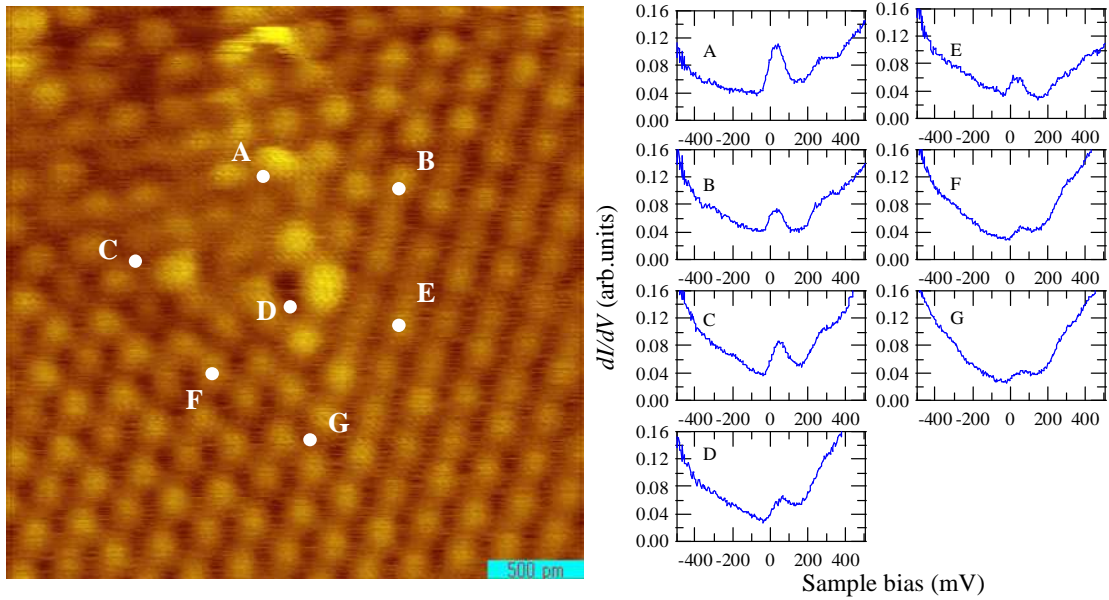

**FIG. S16** STS spectra measured at positions from A to G in the STM image (around Type-A defect in Fig. S10) (Sample bias -500 mV, current 39.9 pA)

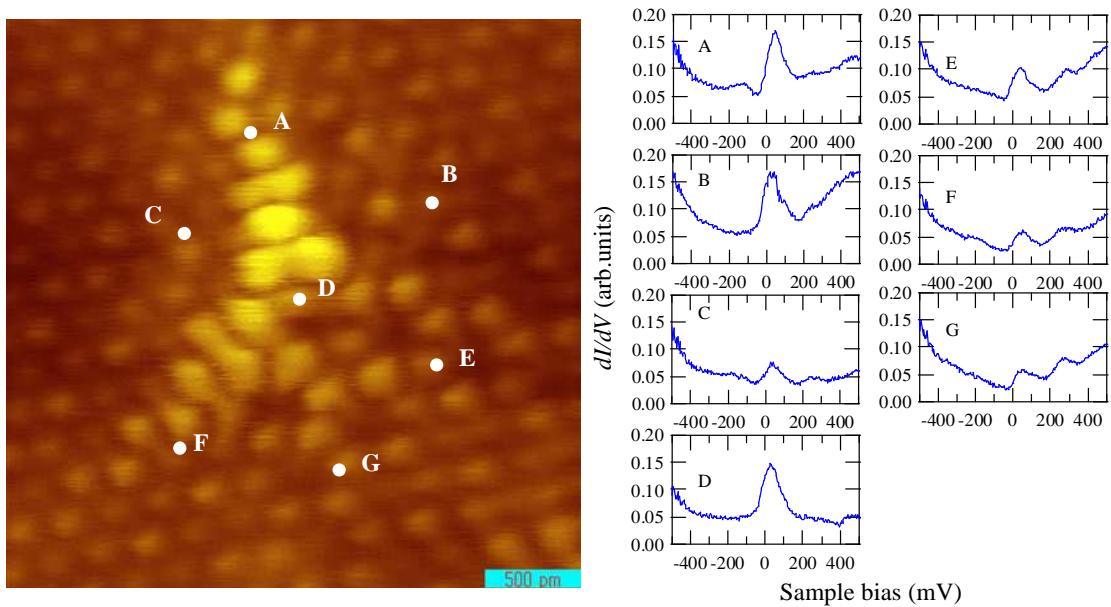

**FIG. S17** STS spectra around Type-B defect in Fig. S1 (Sample bias -500 mV, current 38.9 pA)

Contrary to the cases of Figs. S6-S14, clear position dependence of STS peak intensity is observed in the case of Figs. S16 and S17. Here, two types of defects coexist closely as shown in Fig. S15. The interference of each electronic perturbation may complicate the STS spectrum shape and thus it is not easy to interpret these peaks. As for the STS peak energy, however, all peaks appear at unoccupied states, which is the same STS character of the graphitic-N and single vacancy, though the energy position is not the same, possibly due to their mutual interference.